%% file: sample-sigconf-authordraft.tex
\documentclass[sigconf]{acmart}
\AtBeginDocument{%
  }

\setcopyright{acmlicensed}
\copyrightyear{2025}
\acmYear{2025}
\setcopyright{cc}
\setcctype{by-nc}
\acmConference[SA Technical Communications '25]{SIGGRAPH Asia 2025
Technical Communications}{December 15--18, 2025}{Hong Kong, Hong Kong}
\acmBooktitle{SIGGRAPH Asia 2025 Technical Communications (SA Technical Communications '25), December 15--18, 2025, Hong Kong, Hong Kong}
\acmDOI{10.1145/3757376.3771379}
\acmISBN{979-8-4007-2136-6/2025/12}

\usepackage{subcaption}

\begin{document}

\title[Physics-Based Motion Refinement for Video-Based Capture]{The Last Mile to Production Readiness: Physics-Based Motion Refinement for Video-Based Capture}

\author{Tianxin Tao}
\email{ttao@ea.com}
\orcid{1234-5678-9012}
\affiliation{%
  \department{SEED}
  \institution{Electronic Arts}
  \city{Vancouver}
  \country{Canada}
}

\author{Han Liu}
\email{haliu@ea.com}
\affiliation{%
  \department{SEED}
  \institution{Electronic Arts}
  \city{Redwood City}
  \country{United States}
}

\author{Hung Yu Ling}
\email{hling@ea.com}
\affiliation{%
  \institution{Electronic Arts}
  \city{Vancouver}
  \country{Canada}
}

\renewcommand{\shortauthors}{Tao et al.}

\begin{abstract}
  High-quality motion data underpins games, film, XR, and robotics. Vision-based motion capture tools have made significant progress, offering accessible and visually convincing results, yet often fall short in the final stretch—the last mile—when it comes to physical realism and production readiness, due to various artifacts introduced during capture. In this paper, we summarize key issues through case studies and feedback from professional animators to set a stepping stone for future research in motion cleanup. We then present a physics-based motion refinement framework to bridge the gap, with the goal of reducing labor-intensive manual clean-up and enhancing visual quality and physical realism. Our framework supports both single- and multi-character sequences and can be integrated into animator workflows for further refinement, such as stylizing motions via keyframe editing.
\end{abstract}



\begin{teaserfigure}
    \centering
    \captionsetup{width=1.0\textwidth}
    \includegraphics[width=0.95\linewidth]{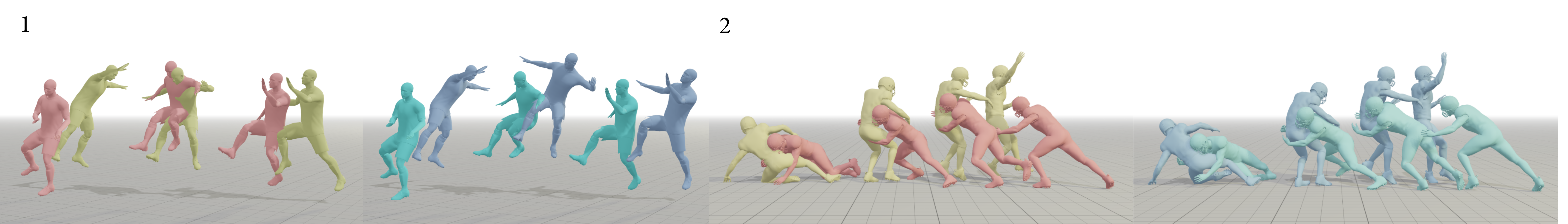}
    \caption{We present a framework for generating physically plausible motions from imperfect vision-based capture data. Two motion clips are presented: 1. torso penetration during header; 2. arm-torso penetration from retargeting. Yellow\&pink characters: raw motions. Teal\&blue characters: cleaned-up results.}
    \label{fig:interpenetration}
\end{teaserfigure}


\maketitle

\input{sections/introduction}
\input{sections/related_work}
\input{sections/problem_statement}

\input{sections/method}
\input{sections/two_character}
\input{sections/results}
\input{sections/discussion_conclusion}


\bibliographystyle{ACM-Reference-Format}
\bibliography{references, additional_references}


\end{document}

%% file: sections/introduction.tex
\section{Introduction}

Motion capture~(Mo-cap) is central to character animation. It has evolved from traditional marker-based and inertia sensor-based systems to recent vision-based techniques, greatly reducing setup effort and enabling scalable motion data collection. Vision-based capture, both in academia~\cite{ye2023decoupling, kanazawa2019learning, TRACE} and in commercial products~\cite{hawkeyeinnovations, tracab, moveai, deepmotion}, now delivers visually appealing results in diverse environments. However, raw vision-captured data still requires manual cleanup for production quality due to subtle but critical artifacts of physical implausibility. In this paper, we first aim to bridge this gap by systematically identifying and categorizing motion artifacts, such as floating, mesh interpenetration, phantom contact, and foot sliding, through feedback and insights gathered from professional animators, noting their persistence even in high-quality commercial pipelines, limiting large-scale deployment of captured motions. To close this "last-mile" gap, we build a physics-based motion refinement framework using deep reinforcement learning~(DRL) that operates on a per-sequence basis, and corrects flawed motion while preserving the integrity and fidelity of the original content. It supports both single- and multi-character scenarios, and introduces key innovations such as artifact-aware evaluation, adaptive termination, and flexible reward shaping.

%% file: sections/related_work.tex
\section{Related Work}


Motion cleanup follows two main paths: physics-based and data-driven. Data-driven methods~\cite{holden2018robust, chen2021mocap} require large, high-quality datasets, which are not always available, so we focus on physics-based approaches. Early work improved quality by simulating marker–character forces~\cite{zordan2003mapping}, and addressed artifacts like ground penetration via sampling-based control~\cite{liu2010sampling}. Subsequent research extended these methods to broader scenarios~\cite{2018-TOG-SFV, Yu:2021:MovingCam}. Recent DRL-based imitation  refines motions from 3D pose estimators~\cite{zhang2023vid2player3d, yuan2021simpoe}, motions involving multi-character interactions~\cite{huang2022neural}, and retargeted motions with contact-graphs~\cite{zhang2023simulation}. \citet{ma2021learning} proposed space-time bounds as an alternative to tracking rewards. 

%% file: sections/problem_statement.tex
\begin{figure}
    \centering
    \includegraphics[width=.7\linewidth]{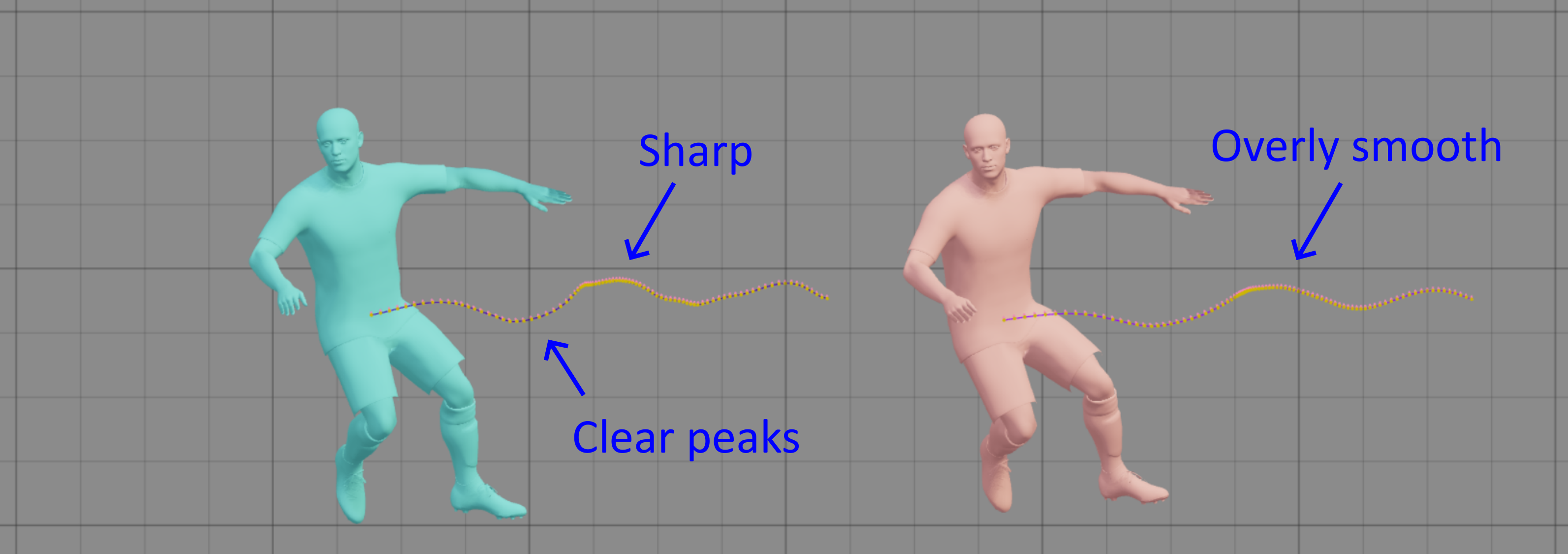}
    \caption{A sharp stop motion showing over-smoothed hip (root) trajectory in the raw input (pink), which lacks a sense of weight. The refined motion (teal) from our method adds realistic root movement.}  
    \label{fig:hip_variation_weightedness}  
\end{figure}

\begin{figure}
    \centering  
    \includegraphics[width=.7\linewidth]{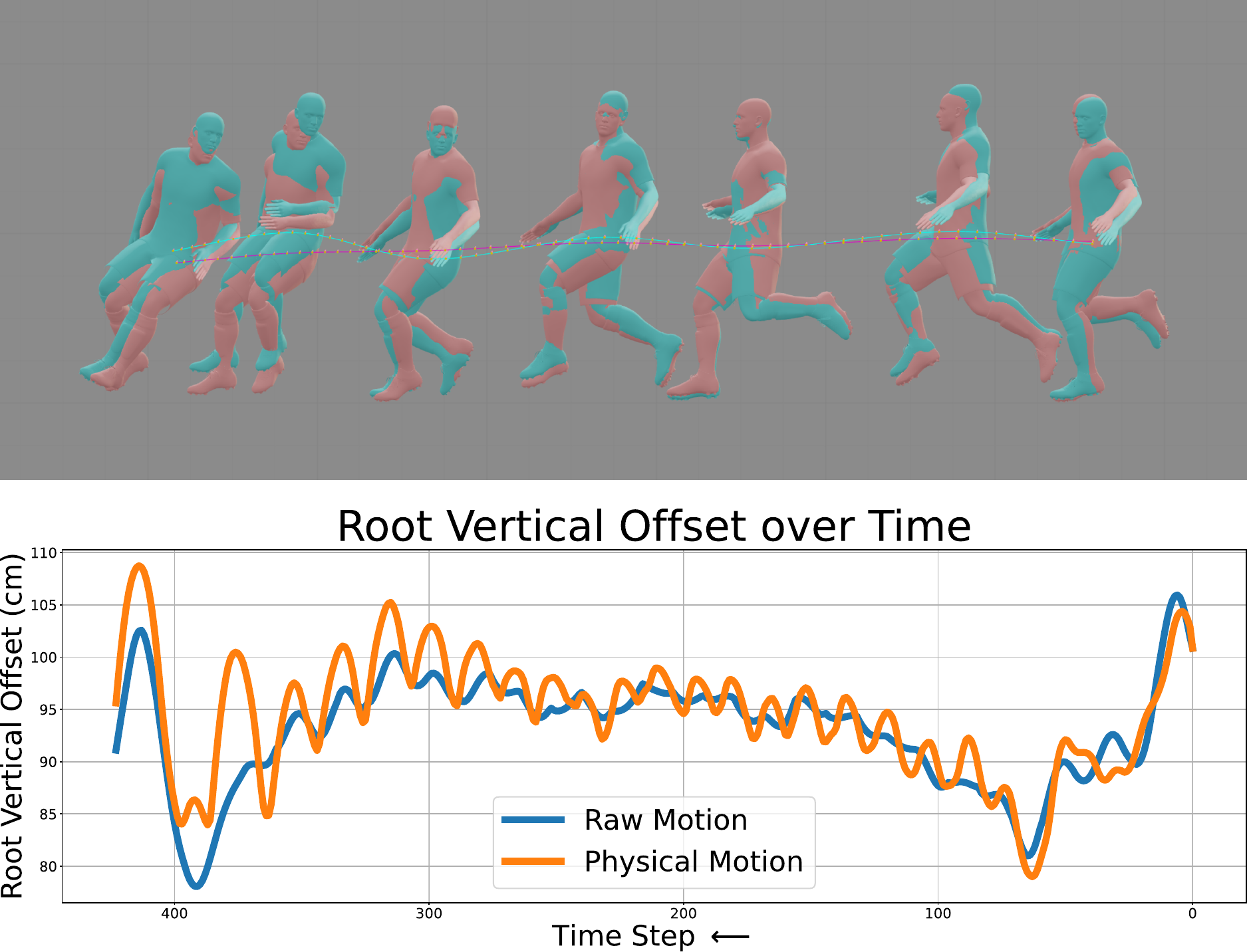}
    \caption{Over-smoothed hip trajectory during locomotion. The plot compares hip height from raw motion (pink) and our physical reconstruction (teal).}  
    \label{fig:hip_variation} 
\end{figure}

\section{Types of Motion Artifacts} \label{sec:problems}

A key motivation for leveraging physics in motion refinement is the observation that many artifacts trace to violations of physical laws. Physical realism is a necessary condition for high-quality motion---any motion that breaks physics principles, such as gravity, friction, or collision constraints, is inherently flawed. We highlight common physics artifacts that impact the quality of captured data. 

\paragraph{Floating and Weightedness.}A typical issue in captured data is the lack of perceived weight. Characters often appear to glide or float, especially during dynamic actions like running or abrupt stops~(see Figure~\ref{fig:hip_variation_weightedness}), despite feet appearing correctly planted. This is usually caused by an over-smoothed root trajectory, which flattens the vertical movement and reduces impact forces, as shown in Figure~\ref{fig:hip_variation}. An over-smooth root trajectory can be easily recognized by expert animators as low-quality motion. Physics-based simulators help resolve this by modeling gravity and inertial forces, allowing characters to exhibit natural and grounded interactions.

\begin{figure}
    \centering  
    \includegraphics[width=.7\linewidth]{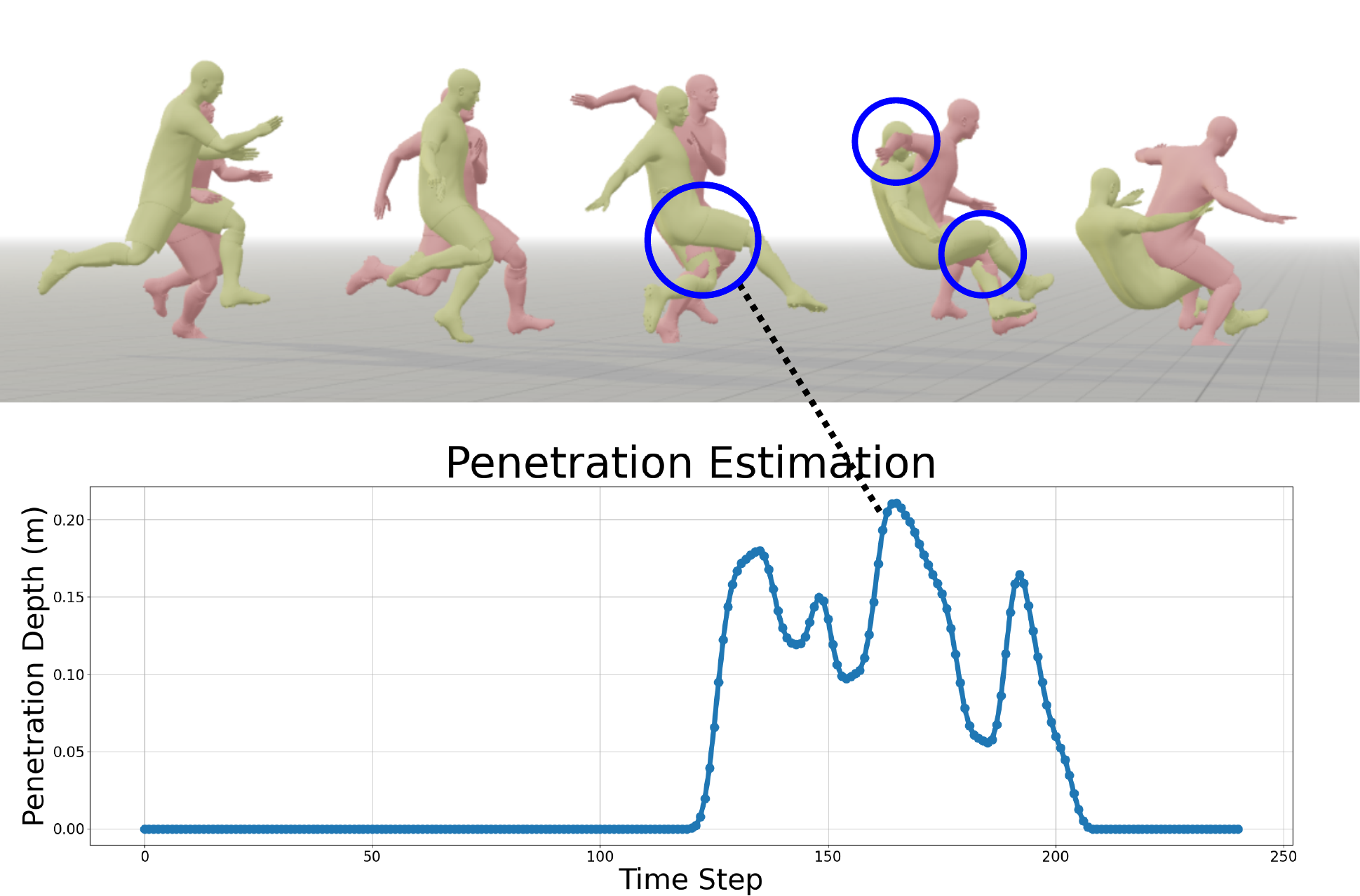}
    \caption{Penetration depth in the motion clip where two characters run into each other. The peaks in the penetration depth are highlighted with blue circles.}  
    \label{fig:penetration}
\end{figure}

\paragraph{Penetration.}Self-penetration and inter-character penetrations are common in both single and multi-character motions. In Figure~\ref{fig:penetration}, we show that these artifacts occur, such as arms that intersect the torso or legs that collide, because pose estimators fail to account for the volumes of physical collisions.
Physics simulators inherently enforce collision constraints at each step, making them a natural tool to address these issues.

\paragraph{Phantom Contact.}When characters interact with objects, exact contact timing is always difficult to capture, which results in phantom contact. This usually stems from incomplete object modeling. For simpler objects, the issue can be addressed by incorporating object motion tracking into the physics-based simulation to enhance contact realism. However, fully modeling all objects remains impractical. As an alternative, we adopt keyframing, a widely used technique familiar to animators, to guide motion cleanup. Animators manually add contact keyframes, and our system automatically refines the motion, reducing artifacts while maintaining realism.

\begin{figure}
    \centering  
    \includegraphics[width=.7\linewidth]{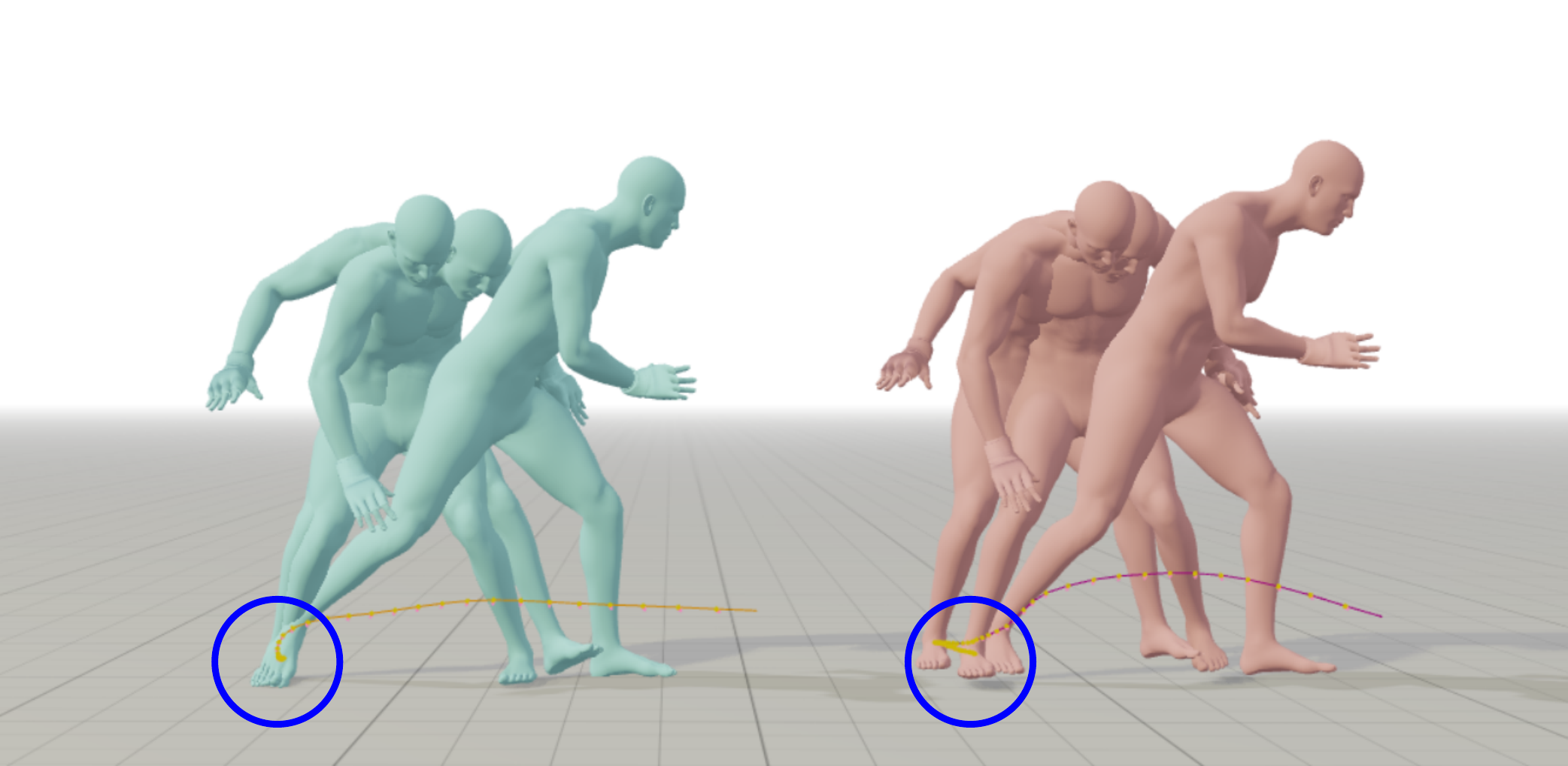}
    \caption{Foot sliding example. The pink character shows sliding artifacts, while the teal version from our method maintains proper foot contact. Blue circle marks a planted right foot, which should remain stationary.}  
    \label{fig:foot_sliding}
\end{figure}

\paragraph{Foot Sliding.}Foot sliding arises from errors in predicting global trajectory and joint positions, especially in the feet, which are sensitive end effectors. In vision-based capture, this leads to visible drifting during ground contact, as illustrated in Figure~\ref{fig:foot_sliding}. Physics-based simulators mitigate this by modeling ground friction to keep feet properly anchored.

%% file: sections/method.tex
\section{System Overview} \label{sec:system_overview}
Our framework addresses the four aforementioned motion artifacts. Floating and foot sliding are naturally mitigated through physics-based motion tracking, while penetration requires additional handling due to the complexity of resolving geometry collisions without compromising motion fidelity. Phantom contact is addressed case by case, depending on the nature of object interaction.

Our framework begins with estimating the severity of artifacts in Mo-cap. From our observations, we have found that measuring penetration serves as an effective indicator of artifact severity. Specifically, we assess self-penetrations to address common issues such as limb clipping, which typically occurs in single characters. For complex interactions involving multiple characters, we measure penetrations between characters to determine the extent of correction required. We then enhance the standard physics-based motion imitation setup, similar to DeepMimic~\cite{peng2018deepmimic}, with key modifications to early termination, reward design, and action space. 

\subsection{Penetration Estimation} \label{sec:penetration_estimation}

Penetration depth intuitively measures the extent of the intersection of objects. Larger depths correspond to more noticeable and unrealistic mesh clipping issues that require correction. Using the articulated humanoid model, we first run collision detection, followed by penetration estimation, specifically, the Gilbert-Johnson-Keerthi~\cite{gilbert1988fast} Algorithm and the Expanding Polytope Algorithm~\cite{van2001proximity}.

In Figure~\ref{fig:penetration}, we illustrate the penetration depths$~\rho_{t}$ in an animation clip involving a side-by-side crash between two characters. At each timestep~$t$, given all the penetration pairs, the penetration depth~$\rho_{t}$ is computed as the maximum value among all body pairs. Penetration depth serves as a reliable indicator of where artifacts exist. It precisely captures common issues such as clipping legs, overlapping limbs, and penetrations during character interactions.

\subsection{Motion Imitation} \label{sec: single_player_imitation}

\paragraph{State and Action.}The state vector~$s_{t}$ contains root information, including root height, root linear velocity, root angular velocity, and root orientation; as well as joint-level attributes, including the joint orientations~$q_{t}$ and joint angular velocities~$\omega_{t}$. In addition, the state further concatenates three future target frames from the reference, represented by the joint positions and linear joint velocities with respect to the character's current root coordinate.

We utilize a proportional-derivative~(PD) controller generating the target joint orientation~$q' = q_{b} + q_{r}$, combining a base orientation~$q_{b}$ and a residual~$q_{r}$ from the policy. While it is common to use the next-frame reference~$\hat{q}_{t+1}$ as ~$q_{b}$, it can mislead the policy when artifacts are present.
Instead, we propose adopting a hybrid PD base by interpolating between the current pose and next reference pose using SLERP:~$q_{b} = \text{SLERP}(q_{t}, \hat{q}_{t+1}, \alpha)$, where~$\alpha \in [0, 1]$ is a one-dimensional scalar predicted by the policy.

\paragraph{Adaptive Termination.}We make several modifications to the motion imitation pipeline to better handle imperfect reference motions. 
We compute the joint position tracking error between the simulated character and the reference in the world coordinate, defined as~$d_{p}^{j} = ||p_{t}^{j} - \hat{p}_{t}^{j}||_{2}$, for joint~$j$. When any~$d_{p}^{j}$ exceeds a threshold, the episode is terminated. While a fixed threshold of 0.5m works well for clean data, it can be overly strict and brittle when the reference motion is corrupted. To address this, we scale~$d_{p}^{j}$ based on the estimated penetration depth~$\rho_{t}$, reducing the penalty in affected frames. Specifically, we relax the position tracking error as:
$d_{p}^{j} = \max(d_{p}^{j} - \kappa * \rho_{t}, 
0),$
where~$\kappa=3.0$ in our implementation. As a result, early termination becomes more forgiving in the presence of artifacts, while still generating physically plausible motions.

\paragraph{Reward.}Our reward function encourages accurate tracking while preserving motion realism. It includes three main terms: joint orientation differences, scaled joint position error~$d_{p}^{j}$, and energy regularization.
The scaled position error~$d_{p}^{j}$ also modulates both the reward and termination condition, allowing the policy to adapt when artifacts appear.
To address phantom contacts, we add a term tracking the motion of interacting dynamic objects. More details are included in the Supplementary Material
~C.

%% file: sections/two_character.tex
\section{Multi-character Motion Cleanup} \label{sec:multiplayer}


\paragraph{Multi-character Control.}We formulate this as a multi-agent DRL problem using Multi-agent PPO~(MAPPO)~\cite{yu2022surprising}. Each character~$m$ has its own policy~$\pi_{m}(s_{t}^{m})$, with actions defined as in the single-character case. A central value function~$V(s^\text{shared}_{t})$ serves as the global critic. Agents receive individual rewards and termination signals, with synchronized resets triggered if any character violates the adaptive termination threshold. We first tried using a single policy to control all characters. Although it worked in some cases, it failed in hard scenarios involving significant penetrations. We found that MAPPO produced more consistent results.

\paragraph{State and Termination Handling.}Each character’s state~$s_{t}^{m}$ includes the its own state~$s_{t}$ and the relative joint positions of other characters~$p^{n}_{t}$, defined in the root frame of~$m$. The shared state~$s^{\text{shared}}_{t}$ is formed by concatenating all~$s_{t}^{m}$. For proper credit assignment during termination, we apply value bootstrapping. If any agent triggers early termination, others receive an updated reward~$r_{t}^{m} = r_{t}^{m} + V(s_{t+1}^{\text{shared}})$ and are also terminated. This ensures consistent~GAE($\lambda$) computation and improves training stability.

%% file: sections/results.tex
\section{Results} \label{sec:results}
The raw motions presented in this section are acquired through commercial vision-based Mo-cap tools—generally high quality but with flaws.
As static images may under-represent the cleanup gains, please refer to the supplementary video for a clearer comparison.
Implementation details can be found in 
Supplementary Material~D.
\subsection{Motion Clean-up}

\begin{figure}
    \centering
    \begin{subfigure}[b]{0.9\linewidth}
        \centering
        \captionsetup{width=0.98\linewidth}
        \includegraphics[width=\linewidth]{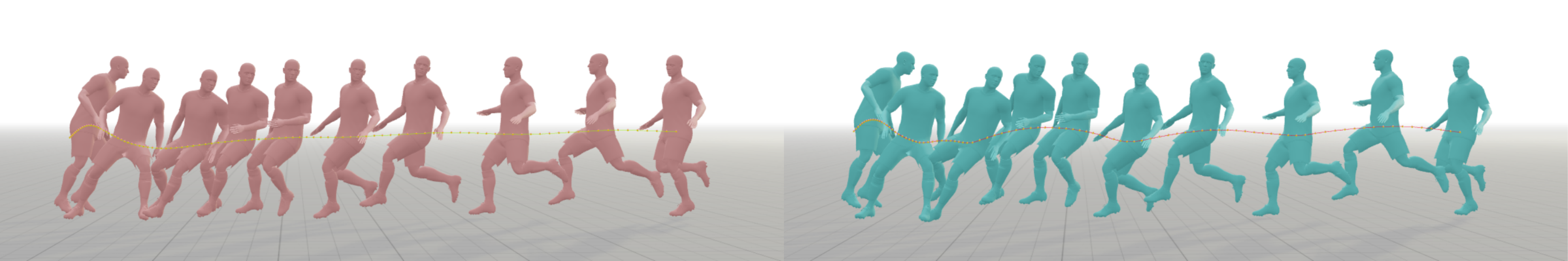}
        \caption{Example demonstrating improved hip movement.}
        \label{fig:sub_clip_3}
    \end{subfigure}
    \begin{subfigure}[b]{0.9\linewidth}
        \centering
        \captionsetup{width=0.98\linewidth}
        \includegraphics[width=\linewidth]{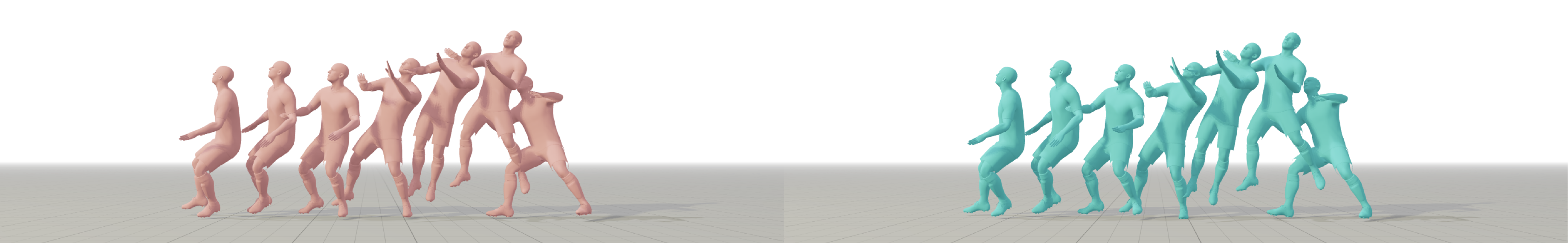}
        \caption{Example demonstrating improved foot slide and enhanced hip motion in the jump.}
        \label{fig:sub_clip_6}
    \end{subfigure}
        \begin{subfigure}[b]{0.9\linewidth}
        \centering
        \captionsetup{width=0.98\linewidth}
        \includegraphics[width=\linewidth]{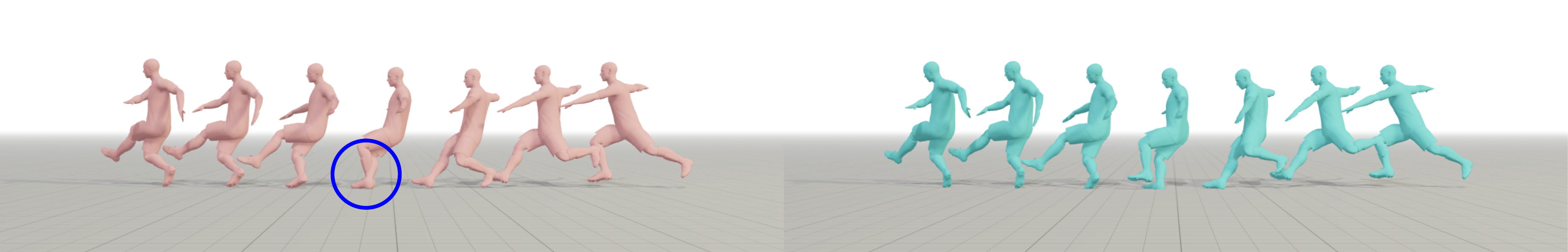}
        \caption{Soccer shot example showing penetration avoidance.}
        \label{fig:sub_clip_7}
    \end{subfigure}
    \begin{subfigure}[b]{0.9\linewidth}
        \centering
        \captionsetup{width=0.98\linewidth}
        \includegraphics[width=\linewidth]{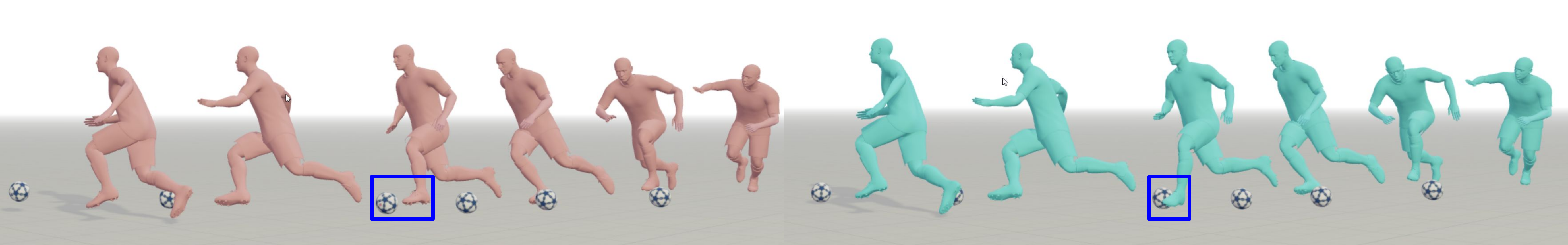}
        \caption{Soccer dribbling example with missing foot contact. The moment of contact is highlighted in the blue box.}
        \label{fig:ball_tracking}
    \end{subfigure}
    \caption{Pink character represents the raw motion, and the teal one represents the motion generated by our framework.}
    \label{fig:results}
\end{figure}

\paragraph{Floating and Weightedness.}Figure~\ref{fig:sub_clip_3} and \ref{fig:sub_clip_6} compare raw and refined motions for enhanced hip movement, with Figure~\ref{fig:sub_clip_3} also plotting root trajectories. This is the same example with vertical root height shown in Figure~\ref{fig:hip_variation}. The refined motion exhibits more sinusoidal root height fluctuations between foot plants, enhancing the perceived sense of weight. In contrast, raw motion often shows flatter trajectories, reducing the sense of gravity and impact. We also list quantitative improvements in 
Supplementary Material~B.

\paragraph{Penetrations.}Mesh clipping is a major issue in raw motion data. As shown in Figure~\ref{fig:sub_clip_7}, our framework resolves self-penetrations such as upper legs intersecting or shins crossing, producing smooth and natural motion. Figure~\ref{fig:interpenetration} further shows that our method also corrects inter-character penetrations during interactions like collisions or hugs, resulting in clean, believable contacts. We report quantitative penetration measurement in 
Supplementary Material~B.

\paragraph{Phantom Contact.}Our framework partially addresses the phantom contact problem. As shown in Figure~\ref{fig:ball_tracking}, a missing contact artifact occurs in a soccer dribbling example. By incorporating soccer ball trajectory tracking, the character successfully executes a timely kick during dribbling. However, scaling the solution to various human-object-interaction problem remains as a challenge.

\paragraph{Foot Sliding.}Our physics-based approach reduces foot sliding by modeling ground friction. As improvements may be difficult to observe in static images, we refer readers to the supplementary video for comparison. A quantitative analysis of foot sliding before and after cleanup is provided in the 
Supplementary Material~B.


\subsection{User-guided Clean-up}

Beyond automatic cleanup, our method also supports animator-guided refinement. In 
Figure~7
in 
Supplementary Material~A
, raw motion with awkward knee and foot orientation during a kick is first partially improved by our framework, but still exhibits unnatural knee bending. An animator then inserts a keyframe for the desired pose and interpolates around it, but this modification degrades kicking dynamics. Applying our framework again produces a smoother, more dynamic motion preferred by the animator. 

Alternatively, animators can edit specific frames directly. As shown in 
Figure~8
in 
Supplementary Material~A, a manual fix to a mispredicted leg pose resulted in unnatural dynamics. Our method restores physical plausibility, improving stability, extension, and landing realism. Please refer to the supplementary video for the visuals of the keyframe technique discussed in the paper.

%% file: sections/discussion_conclusion.tex
\section{Discussion and Conclusion} \label{sec:discussions_conclusions}

While our system produces high-fidelity refinements, it currently requires training a new policy per motion clip, limiting its scalability. Developing a robust, generalizable pretrained policy is an important direction for future work. Additionally, our method assumes the reference motion is structurally sound. When input data is severely corrupted, combining our method with data-driven inpainting or in-betweening approach will improve the overall motion quality. 

In summary, we introduce a physics-based framework for refining physically implausible artifacts in vision-captured motion. Our method improves both single- and multi-character sequences and can be integrated seamlessly into animation workflows. We also analyzed and summarized artifact types that commonly prevent motion from being production-ready, laying the groundwork for future research in motion cleanup.